
\input harvmac

\noblackbox

\Title{\vbox{\baselineskip12pt{\hbox{CTP-TAMU-78/92}}%
{\hbox{hep-th/9212041}}}}
{\vbox{\centerline{Time-Dependent Dilatonic Domain Walls}}}

\centerline{HoSeong ~La\footnote{$^*$}{%
e-mail address: hsla@phys.tamu.edu, hsla@tamphys.bitnet}   }

\bigskip\centerline{Center for Theoretical Physics}
\centerline{Texas A\&M University}
\centerline{College Station, TX 77843-4242, USA}
\vskip 0.7in

Time-dependent domain wall solutions
with infinitesimal thickness are obtained
in the theory of a scalar field coupled to gravity with
the dilaton, i.e. the Jordan-Brans-Dicke gravity.
The value of the dilaton is determined in terms of the Brans-Dicke
parameter $\omega$. In particular, the solutions exist for any $\omega>0$
and as $\omega\to\infty$ we obtain new
solutions in general relativity.
They have horizons whose sizes depend on $\omega$.

\Date{11/92} 
 \noblackbox

\def\tilde{\widetilde}

\def\la{\lambda}
\def\half{{\textstyle{1\over 2}}}
\def\lfr#1#2{{\textstyle{#1\over#2}}} 

\def\e{{\rm e}}
\def\pa{\partial}
\def\mbox#1#2{\vcenter{\hrule \hbox{\vrule height#2in
		\kern#1in \vrule} \hrule}}  

\font\cmss=cmss10 \font\cmsss=cmss10 scaled 833
\def\IZ{\relax\ifmmode\mathchoice
{\hbox{\cmss Z\kern-.4em Z}}{\hbox{\cmss Z\kern-.4em Z}}
{\lower.9pt\hbox{\cmsss Z\kern-.4em Z}}
{\lower1.2pt\hbox{\cmsss Z\kern-.4em Z}}\else{\cmss Z\kern-.4em Z}\fi}

\def\CS{{\cal S}}

\vfill\eject


In the cosmological models for the early universe based on particle physics
and general relativity (GR), it is known that these models admit topological
defects as their classical solutions%
\ref\Vilrev{For reviews, see A. Vilenkin, Phys. Rep. {\bf 121}
(1985) 263.}\ref\KoTu{E.W. Kolb and M.S.
Turner, {\it`` The Early Universe,"} (Addison-Wesley, New York, 1989).}.
These topological defects in principle can be formed
when the universe goes through phase transitions,
although  the chances of actual existence are not necessarily high.
We however anticipate that GR may not be the
ultimate theory to describe the very early universe for which we need a quantum
gravity. Thus we should look for other possibilities how these topological
defects can be formed in other gravitational theories. Besides, the ultimate
theory should not only be able to predict the existence of the existing objects
but also be able to disprove the nonexistence of the nonexisting objects.
In string theory context there have been various approaches, but the complete
results are still elusive\ref\strCos{B.R. Greene, A. Shapere, C. Vafa and S.-T.
Yau, Nucl. Phys. {\bf B337} (1990) 1.}\ref\sDoW{K. Choi and J.E. Kim, Phys.
Rev. Lett. {\bf 55} (1985) 2637;J.A. Casas and  G.G. Ross, Phys.
Lett. {\bf B198} (1987) 461; M. Cvetic, F. Quevedo and S.J. Rey, Phys.
Rev. Lett. {\bf 67} (1991) 1836.}\ref\strMono{J. Harvey
and J. Liu, Phys. Lett. {\bf B268} (1991) 40.}.

In this paper we shall take another possible gravitational theory, namely,
the Jordan-Brans-Dicke (JBD) theory, which is a gravitational
theory with the dilaton in the four-dimensional space-time.
{}From the string theory's point of view, the JBD theory with a specific
Brans-Dicke (BD) parameter is a natural
effective gravitational theory before the dilaton freezes up.
Furthermore, a cosmological  model can also be built based on the JBD
theory\ref\DLaSt{D. La  and P. Steinhardt, Phys. Rev. Lett. {\bf 62} (1989)
376.}. Thus it is
worth while to investigate the existence of topological defects in this
context. Cosmic string
solutions in this theory were previously studied in ref.\ref\GuOr{C.
Gundlach and M.E. Ortiz, Phys. Rev. {\bf D42} (1990) 2521.} and
static thin domain wall solutions have been found in ref.\ref\rDilW{H.S. La,
``Dilatonic Domain Walls," Texas A\&M preprint, CTP-TAMU-52/92
(=hep-ph/9207202) (1992).}. In this paper we shall look for time-dependent
thin domain wall solutions in the JBD theory.

One subtle point in getting a domain wall solution is that solving the
equations
asymptotically is not good enough. Unless it satisfies proper matching
conditions across the wall, it is not a domain wall solution of the system.
As noted in ref.\rDilW, among many asymptotic solutions only one with vanishing
BD parameter survives as a true solution in the static case.
Here we shall adopt the same prescription
for the matching condition at the wall and look for time-dependent solutions.

Let us consider the action for the JBD theory
\eqn\ei{\CS=\int d^4x\sqrt{-\tilde g}\e^{-2\tilde\phi}\left(\tilde{R}
-4\omega\pa_\mu\tilde{\phi}\pa^\mu\tilde{\phi}\right)
+\CS_M[\tilde{g}_{\mu\nu}],}
where $\omega$ is the BD parameter of the theory.
In particular the $\omega=-1$ case corresponds to the action of the dilaton
gravity from string theory\ref\rdilgr{J. Scherk and J.H. Schwarz, Nucl.
Phys. {\bf B81} (1974) 118\semi E.S. Fradkin and A.A. Tseytlin, Nucl. Phys.
{\bf B261} (1985) 1\semi C.G. Callan , D. Friedan, E. Martinec and M.J.
Perry, Nucl. Phys. {\bf B262} (1985) 593.}.
For our purpose we take $\CS_M$ as the action for the real scalar field
with a double well potential:
\eqn\eii{\CS_M=-\half\int d^4x\sqrt{-\tilde g}\left(\pa_\mu\Phi\pa^\mu\Phi
+\la(\Phi^2-\eta^2)^2\right).}
$\CS_M$ in particular has a $\IZ_2$
discrete symmetry $\Phi\to -\Phi$. Note that in the
static case, this discrete symmetry is equivalent to the time reversal
symmetry. This is why the domain wall structure is related to the CP
phases\ref\TDL{T.D. Lee, Phys. Rep. {\bf 9} (1974) 143.}\ref\rZel{Ya.B.
Zel'dovich, I.Yu. Kobzarev, and L.B. Okun, JETP {\bf 40} (1975) 1.}.
Here we shall take $\Phi$ to be static, but the metric tensor can depend on
time. In principle one can generalize to the case in which $\Phi$ depends on
time. The result however should be the same in the thin wall case
if we assume the position of the wall does not change.

For convenience we redefine the variables as
\eqn\eiii{\omega={1\over 4\beta^2}-{3\over 2},\ \
\e^{-2\tilde\phi}={1\over 16\pi G}\e^{-2\beta\phi},\ \
\tilde{g}_{\mu\nu}=\e^{2\beta\phi} g_{\mu\nu},}
then the action now becomes
\eqn\esa{\CS={1\over 16\pi G}\int d^4x\sqrt{-g}(R-\pa_\mu\phi\pa^\mu\phi)
+\CS_M[\e^{2\beta\phi}g_{\mu\nu}].}
Setting $\phi=0$, this action reduces to that of a real scalar field
coupled to the Einstein gravity, in which  case a time-dependent thin
domain wall solution is
known to exist\ref\ViIpS{A.~Vilenkin, Phys. Lett. {\bf 133B} (1983) 177;
J.~Ipser and P.~Sikivie, Phys. Rev. {\bf D30} (1984)712.}.

In our case although $\tilde{g}_{\mu\nu}$ is a physical metric of the
space-time, eq.\esa\ is very convenient for practical purposes.
Although $\tilde\phi$ is the actual dilaton, we will call $\phi$ a dilaton too
since they are related by a simple field redefinition.
We shall also assume that $\phi$ does not depend on time.

\def\sqg{\sqrt{-g}}
\def\mn{\mu\nu}
\def\tph{\tilde{\phi}}

By varying eq.\esa\ with respect to the new metric $g_{\mu\nu}$, we obtain
\eqn\emet{R_{\mu\nu}=\pa_\mu\phi\pa_\nu\phi+8\pi G(T_{\mu\nu}-\half
g_{\mu\nu}T),}
where the ``energy-momentum'' tensor is given by
\eqn\eten{T_{\mu\nu}=-{2\over\sqrt{-g}}{\delta\CS_M\over\delta g^{\mu\nu}}
=-\half\left[\e^{2\beta\phi}g_{\mu\nu}g^{\alpha\beta}\pa_\alpha\Phi\pa_\beta\Phi
-2\e^{2\beta\phi}\pa_\mu\Phi\pa_\nu\Phi+g_{\mu\nu}\e^{4\beta\phi}\la
(\Phi^2-\eta^2)^2\right].}
Note that $T_{\mu\nu}$ is not covariantly
conserved due to the dilatonic contribution.
The physical energy-momentum tensor $\tilde{T}^{{\rm
matter}}_{\mn}+\tilde{T}^{\tilde{\phi}}_{\mn}$
satisfies the gravitational equation of motion for JBD theory\ref\Wein{S.
Weinberg, ``{\it Gravitation and Cosmology}," (Wiley, New York, 1972).},
\eqn\jbeom{\tilde{R}_{\mn}-\half\tilde{g}_{\mn}\tilde{R}
=8\pi\e^{2\tilde{\phi}} \left(\tilde{T}^{{\rm
matter}}_{\mn}+\tilde{T}^{\tilde{\phi}}_{\mn}\right),}
where $\nabla^\mu\tilde{T}^{{\rm matter}}_{\mn}=0$ and
\eqn\dilem{\eqalign{
\tilde{T}^{\tilde{\phi}}_{\mn}=&
{1\over 8\pi}\e^{-2\tph}\left(2\pa_\mu\pa_\nu\tph
+2\tilde{\Gamma}^{\alpha}_{\mn}\pa_\alpha\tph+\half\e^{2\tph}\tilde{g}_{\mn}
\tilde{\mbox{.1}{.1}}^2\e^{-2\tph}\right)\cr
&-{\omega\over 2\pi}\e^{-2\tph}
\left(\pa_\mu\tph\pa_\nu\tph-\half\tilde{g}_{\mn}\pa^\alpha\tph\pa_\alpha\tph
\right) .\cr}}
The second term of eq.\dilem\ is proportional to $\omega$ so that it
vanishes if $\omega=0$, but the first term is independent of $\omega$.

The field equations for the dilaton $\phi$ is
\eqn\edil{{\mbox{.1}{.1}}^2\phi={1\over\sqg}\pa_\mu(\sqg g^{\mn}\pa_\nu)\phi=
-8\pi G\beta T}
and the matter scalar field satisfies
\eqn\ehig{\pa_\mu(\sqg \e^{2\beta\phi} g^{\mn}\pa_\nu)\Phi
-2\la\sqg\e^{4\beta\phi}\Phi(\Phi^2-\eta^2)=0.}

In general, domain wall solutions are obtained in theories where a discrete
symmetry is spontaneously broken. Note that the action eq.\eii\ for the
matter field $\Phi$ has a discrete symmetry $\Phi\to -\Phi$ so that we can
look for domain walls, when this symmetry is spontaneously broken.
In the case where domain walls have infinitesimal
thickness, we can approximate the wanted scalar field to behave as
$$\Phi(z)=\cases{\eta & if $z>0$;\cr -\eta& if $z<0$.\cr}$$
Then we are interested in the the domain wall which
 separates a space of the $\Phi=\eta$ phase from a space of
the $\Phi=-\eta$ phase.
Such an approximation is  in fact reasonable for the cases where the Compton
wavelength of the test particle is much longer than the thickness of the
wall.

For the static cases we used the following {\it ansatz}
for domain wall solutions\ref\rTa{A.H. Taub, Phys. Rev. {\bf 103} (1956) 454.}%
\ref\doref{A. Vilenkin, Phys. Rev. {\bf D23} (1981) 852.}\rDilW:
$${ds^2=A(|z|)(-dt^2+dz^2) +B(|z|)(dx^2+dy^2).}$$
Note that we have required the reflection symmetry between each side of
 the wall, which is an
infinite plane perpendicular to the $z$-direction at $z=0$.
Now for time-dependent domain wall solutions we shall try
\eqn\tdoastz{ds^2=-A(|z|)dt^2+C(|z|)dz^2 +B(|z|)U(t)(dx^2+dy^2).}
Again we have the reflection symmetry. The time dependent part of the metric is
added in $g_{xx}=g_{yy}$.
This ansatz in fact is a generalized version of the time-dependent domain wall
solutions in GR, in which case  $C=1$\ViIpS.

Now we need a prescription to take care of the matching conditions across the
wall. Strictly speaking $|z|$ is not analytic at $z=0$. However from physics'
point of view this merely is an approximation due to the thin wall assumption.
One should treat it as a limit case of analytic function. One way to
introduce a reasonable analytic property for $|z|$ is to use a step function.
Thus we can use the following
prescription for $|z|$ to avoid such a difficulty\rDilW :
\eqn\ezpre{|z|=z[\theta(z)-\theta(-z)],}
  where
$\theta(z)$ is a step function defined by $\theta(z)=1$ for $z\geq 0$,
$\theta(z)=0$ for $z<0$. Then
$\pa_z|z|=[\theta(z)-\theta(-z)]+2z\delta(z)$.
If we promise that $\pa_z|z|$ shall be multiplied
with some function of $z$ that does not have a pole at $z=0$, we
can safely use an identification $\pa_z|z|\equiv\theta(z)-\theta(-z)$.
Similarly,
$\pa_z^2|z|\equiv 2\delta(z)$. The reason we try to be careful about such
analyticity is to check the consistency of the solutions at the wall, which
turns out to be important to
provide interesting constraints on the solutions.
Again we would like to emphasize that this should be regarded as an
approximation.

Using this ansatz, eq.\tdoastz, we find
\eqn\ertz{R_{tz}=\half{\dot{U}\over U}\left({A'\over A} -{B'\over B}\right),}
where the prime and the dot denote $\pa_z$ and $\pa_t$ respectively.
The corresponding field equation $R_{tz}=0$ leads to
\eqn\euv{A=B.}
With this identification the field equations now become
\eqna\comfi
$$\eqalignno{
R_{tt} &=\lfr{1}{4}{A'^2\over AC} -\lfr{1}{4} {A'C'\over C^2}
+\half {A''\over C} +\half {{\dot U}^2\over U^2} -{{\ddot U}\over U}
=-4\pi G\la A\e^{4\beta\phi}\left(\Phi^2-\eta^2\right)^2, &\comfi a\cr
R_{zz} &=\lfr{3}{4}{A'^2\over A^2} +\lfr{3}{4} {A'C'\over AC} -\lfr{3}{2}
{A''\over A}
=\!\phi'^2\!\!+\!8\pi G\!\left(
\e^{2\beta\phi}\Phi'^2\!+\half\la C\e^{4\beta\phi}
(\Phi^2\!\!-\!\eta^2)^2\right)\!\!,
&\comfi b\cr
R_{xx}&=R_{yy} &\cr
&= U\!\left(
-\lfr{1}{4}{A'^2\over AC} +\lfr{1}{4} {A'C'\over C^2}
-\half {A''\over C} +\half{{\ddot U}\over U} \right)
\!=4\pi G\la A U\e^{4\beta\phi}\left(\Phi^2-\eta^2\right)^2,
\ \ \ \ \ \  &\comfi c\cr
&{1\over A^{3/2} C^{1/2}}\left({A^{3/2}\over C^{1/2}}\phi'\right)'
=8\pi G\beta\left(\e^{2\beta\phi}{1\over C}\Phi'^2 +2\la
\e^{4\beta\phi}(\Phi^2-\eta^2)^2\right)\!\!, &\comfi d\cr
&\left({A^{3/2}\over C^{1/2}}\e^{2\beta\phi}\Phi'\right)' =2\la
A^{3/2}C^{1/2} \e^{4\beta\phi}\Phi(\Phi^2-\eta^2). &\comfi e\cr}$$

For thin walls we have $\Phi^2=\eta^2$ and $\Phi'=0$ away from $z=0$ so
that we shall first solve the above equations away from $z=0$, then shall check
the consistency at the wall.
Solving eqs.\comfi{a,c}\ for $z\neq 0$, we obtain
\eqn\solut{U(t)=\e^{\kappa t},}
and from eq.\comfi{d} we obtain
\eqn\solphi{\phi'=\alpha{C^{1/2}\over A^{3/2}},}
where $\kappa, \alpha$ are constants yet to be determined.

Using these in eqs.\comfi{a-c}, we can determine
\eqn\solc{C={AA'^2\over \left(\kappa^2 A^2+{2\over 3}\alpha^2\right)}\ \ .}
With such $C$ eq.\solphi\ now can be integrated as
\eqn\esphis{\phi=\half\sqrt{\lfr{3}{2}}\ln\left({
\sqrt{1+{3\kappa^2\over 2\alpha^2}A^2}-1\over
\sqrt{1+{3\kappa^2\over 2\alpha^2}A^2}+1}\right),}
where we have taken $A(0)=1, C(0)=1,$ and accordingly $\phi(0)$.

Since this eq.\solc\ satisfies eqs.\comfi{a-c}\ identically away from $z=0$,
there is no other constraint on the function $A$ so that for any $A$ we can
obtain an asymptotic solution. This is rather intriguing, although
some of them might fail to satisfy the matching conditions across the wall.
Also some proper asymptotic boundary conditions will distinguish them.
For practical purposes, it is more convenient to choose $C$
and solve for $A$ to check the matching conditions across the wall.

If we assume $C=1$, then for $\alpha=0$ we reproduce the known solutions in the
general relativity case\ViIpS, which can be shown easily from eq\solc.
For $\alpha\neq 0$ there are other solutions but
the indefinite integral to obtain $A$ cannot be performed analytically.

\def\factak{\sqrt{1+{2\alpha^2\over 3\kappa^2}}}
\def\cosh{{\rm cosh}}
\def\sinh{{\rm sinh}}
{}From now on we shall concentrate on one particularly interesting case of
$C=A$
and leave the rest as future exercises.
For $C=A$ eq.\comfi{a} is precisely the same as eq.\comfi{c}.
Now eq.\solc\ can be integrated to yield
\eqn\esola{A=p\e^{\kappa z}+q\e^{-\kappa z},}
where
\eqn\solacon{4pq\kappa^2+\lfr{2}{3}\alpha^2=0.}
With the required boundary condition, $A(0)=1$, {\it i.e.} $p+q=1$,
we can explicitly determine $p,q$ as
\eqn\solpq{p=\half\left(1-\sqrt{1+{2\alpha^2\over 3\kappa^2}}\right),\ \
q=\half\left(1+\sqrt{1+{2\alpha^2\over 3\kappa^2}}\right).}
Thus we have
\eqn\fsola{C=A=B=\cosh\kappa |z|-{\factak}\sinh\kappa |z|,}
and from eq.\esphis, we get
\eqn\phisol{\phi=\sqrt{\lfr{3}{2}}
\ln\left|{1-\sqrt{\lfr{3}{2}}{\kappa\over\alpha}\left(1-\factak\right)
\e^{\kappa |z|}\over
1+\sqrt{\lfr{3}{2}}{\kappa\over\alpha}\left(1-\factak\right)\e^{\kappa |z|}}
\right|.}
$\kappa$ shall be determined in terms of the energy density so that we are left
with a free parameter $\alpha$. Thus
these are one parameter family of asymptotic solutions.
But as we mentioned before, not all of them are true domain wall solutions.

Now let us check the consistency of the solution at the wall.
Using the analytic property eq.\ezpre\ we prescribed,
For small $\kappa$ eq.\comfi{a,c}\ reduces to
\eqn\corchi{\kappa \factak\delta(z)= 4\pi G\la\left(\Phi^2-\eta^2\right)^2, }
which leads to $\kappa>0$. Later we shall find out that this small $\kappa$
assumption is indeed related to the weak gravitational field approximation
in the sense that the energy density is small.
Similarly,  eq.\comfi{b}\ leads to
\eqn\ecchi{\kappa \factak\delta(z)= 4\pi G(\Phi')^2. }
and eq.\comfi{d}\ reduces to
\eqn\eI{\left({\alpha\over \beta}-2\kappa\factak\right)\delta(z)
=4\pi G(\Phi')^2.}

The consistency of eqs.\ecchi\eI\ identifies the coefficients of the
delta-function so that we can determine $\alpha$ in terms of $\beta$ as
\eqn\abrel{\alpha={3\beta\kappa\over\sqrt{1-6\beta^2}}.}
Thus the solution exists only for $\beta^2<1/6$, i.e. $\omega>0$.

Note that $\kappa$ and $\eta$ have mass dimensions and the
Newton's gravitational constant $G$ has inverse mass square dimension, while
$\la$ is a dimensionless coupling constant. Using a dimensional analysis for
a possible thick wall, if we have
$G\la\eta^4\gg {\kappa}{\eta\over\sqrt\la}$ and $\Phi(z=0)=0$,
the LHS of eqs.\corchi\ecchi\ are effectively
comparable to the RHS by smearing out the delta function
Thus this is a good approximate solution and
the condition in fact corresponds to the weak gravitational field limit.

Finally, $\kappa$ can be determined from the ``energy''
density as follows:
Using eqs.\corchi -\eI\ we can compute the ``energy-momentum'' tensor eq.\eten\
as
\eqn\edisen{T_{\mu\nu}={\kappa\over 4\pi G\sqrt{1-6\beta^2}}\delta (z)
{\rm diag} (1, -1, -1, 0),\ \ \ \ \kappa>0.}
As we promised, the small $\kappa$ implies the small energy density.
The corresponding physical energy-momentum tensor can be computed from
eq.\jbeom.

Here we would like to call the readers attention to the fact that we have
differential equations with the Dirac delta-function. Some may find that this
is unreasonable because after all the Dirac delta-function is not a function
but a distribution. But this is not completely unreasonable in field theory
when we often need to be careful about the analyticity. The main intention is
not to solve the differential equations in question
but to check the consistency between
equations. In this sense this is a sufficiently good approximation.
In fact one can be more careful about this situation and can introduce
distributional energy-momentum tensor in terms of delta-function from the
beginning\ref\rIsGer{W. Israel, Nuo. Cim. {\bf 44B} (1966) 1;
R. Geroch and J. Traschen, Phys. Rev. {\bf D36} (1987) 1017.}.
The result however is more or less equivalent because we also have
derived the distributional ``energy-momentum'' tensor using our prescription
eq.\ezpre. We can also further clarify the
result by introducing infinitesimal thickness of the wall and taking
approximation around $z=0$, although we cannot determine the shape of the
solution within this thickness exactly.

Now the physical metric can be obtained by multiplying the conformal factor
(see eq.\eiii) as $d\tilde{s}^2=\e^{2\beta\phi}ds^2$ so that we obtain
\eqn\resone{d\tilde{s}^2=\e^{2\beta\phi} A(|z|)
\left(-dt^2+dz^2+\e^{\kappa t}(dx^2+dy^2)\right),}
where
$$\eqalign{\e^{2\beta\phi}&=
\left|
{\sqrt{6}\beta-\left(1-\sqrt{1-6\beta^2}\right)
\e^{\kappa |z|}\over
\sqrt{6}\beta+\left(1-\sqrt{1-6\beta^2}\right)
\e^{\kappa |z|}
}\right|^{\sqrt{6} \beta},\cr
A(|z|)&=\left(\cosh \kappa |z|-{1\over\sqrt{1-6\beta^2}}
\sinh\kappa |z|\right).\cr}
$$
The value of the dilaton $\tilde\phi$ can be obtained  in terms of $\beta$,
which in turn related to the BD parameter $\omega$ in eq.\eiii, as
\eqn\vdilv{\tilde\phi=\beta\phi+\half\ln 16\pi G.}
In particular, at the wall we have
$$\tilde\phi(0) = \sqrt{\lfr{3}{2}}\beta\ln\sqrt{{1-{\sqrt 6}\beta\over
1+{\sqrt 6}\beta}} +\half\ln 16\pi G.$$
Note that
as $\omega\to\infty$, i.e. $\beta\to 0$, we have new solutions in general
relativity
\eqn\enegrso{A(|z|)=\e^{-\kappa |z|}.}

For a given BD parameter $\omega>0$ we have found time-dependent thin domain
wall solutions in the JBD theory. For $\omega=0$, it is known that stable
static thin domain wall solution exists as proven by the author\rDilW.
This in particular also proves that there is no thin domain wall solutions
for $\omega<0$, i.e. in the 4-d analogue of the dilaton gravity.

As in the GR case,
the time-dependent domain wall we have obtained also has a horizon at
$|z|={1\over \kappa}\tanh^{-1}\sqrt{1-6\beta^2}$. As $\omega$ increases (i.e.
$\beta\to 0$.), the
size of the horizon also increases. As $\omega\to 0$ (i.e. $\beta^2\to 1/6$.),
the horizon shrinks and eventually disappears.
Note that the signature of the metric changes behond the horizon. Thus we
should regard the region within thee horizon also as where our choice of the
coordinate system is valid. This however is not unusual. As is known eeeeveen
in GR case, the metric of time-dependent thin walls become singular as $|z| \to
\infty$.

Although we have shown only one case explicitly, there are enormous amount of
possible solutions according to eeeq.\solc.
At this moment we do not understand why there are so many
solutions. We however
expect that there should be some classification scheme, or some
arguments to exclude some of them.
Also it will be interesting to search for solutions which behave well outside
the horizon.

Much work is needed to understand the structure of all other possible
solutions and
what kind of cosmological implications these
dilatonic domain wall solutions have.

\bigbreak\bigskip\bigskip\centerline{{\bf Acknowledgements}}\nobreak

\par\vskip.3truein

The author thanks S. Fulling for allowing him to access to the MathTensor.
This work was supported in part by NSF grant PHY89-07887 and World Laboratory.


%
\listrefs
\vfill\eject
\bye